\documentclass[bm]{epl}
\usepackage{epsfig}
\begin{document}

\title{Quantum Irreversibility of Energy Spreading}
\shorttitle{Quantum Irreversibility}

\author{Tsampikos Kottos\inst{1}, and Doron Cohen\inst{2}}
\institute{
  \inst{1} \mbox{Max-Planck-Institut f\"ur Str\"omungsforschung,
            Bunsenstra\ss e 10, D-37073 G\"ottingen, Germany} \\
  \inst{2} \mbox{Department of Physics, Ben-Gurion University,
           Beer-Sheva 84105, Israel}
}

\maketitle

\begin{abstract}
The analysis of dissipation and dephasing in driven mesoscopic devices requires a
distinction between two notions of quantum irreversibility. One ("Loschmidt echo")
is related to "time reversal", while the other is related to "driving reversal".
In the latter context we define the time of maximum return ("compensation time")
which generalizes the notion of "echo time".
Non-perturbative features manifest themselves in the energy spreading process.
This is demonstrated for the prototype random-matrix Wigner model, where  the
compensation time and the system response exhibit a non-universal scaling behavior.
\end{abstract}


Quantum Irreversibility (QI) \cite{peres} is a subject of a recent intensive
research activity due to its relevance to quantum computing \cite{quantcomp}.
It turns out that the analysis of dissipation and dephasing in driven mesoscopic
devices requires a distinction between two different notions of "irreversibility".
One is based on the "piston model" paradigm (PMP), while the other is based
on the "ice cube in a cup of hot water" paradigm (ICP). The latter notion has
been adopted by the recent literature \cite{peres,quantcomp}, and it is related
to the studies of dephasing due to the interaction with chaotic degrees of
freedom \cite{wld}. The PMP on the other hand has direct relevance to our
recent studies \cite{rsp} of the (non-perturbative) response of driven chaotic
mesoscopic systems.

In the PMP case we say that a process is reversible if it is possible to "undo"
it.  Consider a gas inside a cylinder with a piston. Let us shift the piston inside.
Due to the compression the gas is heated up. Can we undo the "heating" simply by
shifting the piston outside, back to its original position? If the answer is "yes",
as in the case of strictly adiabatic process, then the process is said to be
reversible.
In the ICP case we consider the melting process of an ice cube. Let
us assume that after some time we reverse the velocities of all the molecules. If
the external conditions are kept strictly the same, we expect the ice-cube to
re-emerge out of the water. In practice the external conditions (fields) are not
exactly the same, and as a result we have what looks like irreversibility.

It is also essential to define precisely what is the {\em measure} for QI.
The prevailing possibility is to take the survival probability as
a measure \cite{peres}. Another possibility is to take the energy spreading
as a measure. (The energy spreading is defined as the square root of the
energy variance.) The latter definition goes well with the PMP, and it has
a well defined classical limit.

Thus the ICP is related to "time reversal" experiments, while the PMP is related
to "driving reversal" experiments. Later we are going to put the focus on PMP irreversibility.
A thorough understanding of the one-period driving reversal scenario is both important
by itself, and also constitutes a bridge towards a theory for the response to periodic driving \cite{rsp}.
It should be realized that irreversibility in the PMP sense implies {\em irreversible} diffusion
in energy space. This diffusion leads to {\em irreversible} energy absorption ("dissipation")
in periodically driven mesoscopic devices \cite{vrn}.
One purpose of this Letter is to sharpen the notion "irreversible diffusion".

The structure of this Letter is as follows: First we discuss the mathematical
formulation of the ICP and the PMP. Then we focus on QI in the PMP sense, which
has not been done in the past as far as we know. In this context we define the
time of maximum return ("compensation time"), which constitutes a generalization
of the notion of "echo time".
The specific model that we study is the well known Wigner model \cite{wigner}.
We shall explain that this is the simplest model that contains
the main ingredients of a physical problem. We define the notions of
perturbative/non-perturbative regimes and in particular we explain the adiabatic/sudden limits.
The nature of QI in the different dynamical regimes is found to be distinct. The
theoretical considerations are supported by the analysis of extensive numerical
simulations. The new insights regarding the analysis of wavepacket dynamics in
energy space \cite{wbr} are summarized in the last paragraph.

Consider a system whose evolution is governed by the chaotic Hamiltonian
${\cal H}={\cal H}(Q_{\alpha},P_{\alpha};x)$, where $(Q_{\alpha},P_{\alpha})$ is a set
of canonical coordinates, and $x=x(t)$  with $0<t<T/2$ is a time dependent parameter.
The corresponding evolution operator will be denoted by~$U[x]$. In the ICP case
the parameter $x$ represents the external fields. The physical conditions that
correspond to the "forward" evolution are represented by $x=x_{\rm {A}}(t)$,
while the physical conditions that correspond to the "backward" evolution are
the time-reversal of $x=x_{\rm {B}}(t)$. If $x_{\rm {B}}(t)=x_{\rm {A}}(t)$
then we have complete reversibility, otherwise we can define a generalized fidelity
amplitude as $F(T)=\langle \Psi_0 |U[x_{\rm {B}}]^{-1} U[x_{\rm {A}}] | \Psi_0
\rangle$, where $\Psi_0$ is the initial preparation. The generalized fidelity
amplitude $F(T)$ is in  fact \cite{wld} the Feynman-Vernon influence functional,
and $|F(t)| \le 1$ is the so called dephasing factor. For constant $x_{\rm {A}}$
and $x_{\rm {B}}$ we can write $U[x_{\rm {A}}]=\exp(-iT{\cal H}_{\rm {A}}/2\hbar)$
and $U[x_{\rm {B}}]=\exp(-iT{\cal H}_{\rm {B}}/2\hbar)$, leading to the currently
prevailing definition of the fidelity, also know as Loschmidt echo, as a measure
for quantum irreversibility \cite{peres}.

To be more specific we assume that the variations of the external fields are
classically small, so that we can linearize the Hamiltonian as follows:
${\cal H} = {\cal H}_0 + x{\cal W}$.
Then, for constant perturbation $x=\pm\epsilon$, we can write ${\cal H}_{\rm {A}}=
{\cal H}_0+\epsilon{\cal W}$ and ${\cal H}_{\rm {B}}={\cal H}_0-\epsilon{\cal W}$,
where $\epsilon$ is the amplitude of the perturbation.  Note that the prevailing
definition of the fidelity ${\cal P}_{\rm {LE}}(T)=|F(T)|^2$, also known as Loschmidt
echo, takes the form
\begin{eqnarray}
\label{e2}
{\cal P}_{\rm {LE}}(T)
= |\langle \Psi_0|\mbox{e}^{-i\frac{T}{2\hbar}(-{\cal H}_0{+}\epsilon{\cal W})}
\mbox{e}^{-i\frac{T}{2\hbar}({\cal H}_0{+}\epsilon{\cal W})}|\Psi_0\rangle|^2
\end{eqnarray}

In complete analogy with the ICP, we can define
${\cal P}(T) = |\langle \Psi_0 |U[x_{\rm {B}}(\mbox{rev})]
U[x_{\rm {A}}] | \Psi_0 \rangle|^2$.
Here $x_{\rm {B}}(\mbox{rev})$ corresponds to the reversed
driving process, eg pushing the piston back to its initial
location. For constant $x_{\rm {A}}$ and $x_{\rm {B}}$ we can
write $U[x_{\rm {A}}]=\exp(-i\frac{T}{2}{\cal H}_{\rm {A}})$ and
$U[x_{\rm {B}}(\mbox{rev})]=\exp(-i\frac{T}{2}{\cal H}_{\rm {B}})$,
and for the linearized Hamiltonian we get
\begin{eqnarray} \label{e3}
{\cal P}(T)=|\langle \Psi_0|\mbox{e}^{-i\frac{T}{2\hbar}({\cal H}_0{-}\epsilon{\cal W})}
\mbox{e}^{-i\frac{T}{2\hbar}({\cal H}_0{+}\epsilon{\cal W})}|\Psi_0\rangle|^2
\end{eqnarray}
If we take the PMP literally, then the parameter $\epsilon$
represents either a small {\em displacement} of a piston,
or its {\em velocity}. The latter interpretation emerges
if we write the Schr\"odinger equation in the adiabatic basis
(See eg Eq.~(55) of Ref.\cite{vrn}).
Similarly, in case of an Aharonov-Bohm ring system,
the parameter $\epsilon$ represents either the magnetic flux
through a ring, or else it may represent
the electro-motive-force (measured in Volts).
The latter, by Faraday law, is the time derivative of the former.
As a matter of terminology, note that if $\epsilon$ is
interpreted as "velocity", then what we call
below "standard perturbative regime" is in fact
the "adiabatic regime" (small $\epsilon$).
Later we also discuss the notion of "sudden regime" (large $\epsilon$).

At first sight the two definitions of ${\cal P}(T)$ look very similar:
it is as if the role of ${\cal H}_0$ and ${\cal W}$ is interchanged.
Also one may (wrongly) think that there is some kind of symmetry
between the "adiabatic limit"  ($\epsilon\rightarrow0$),
and the opposing "sudden limit"  ($\epsilon\rightarrow\infty$).
This superficial "algebraic" symmetry of the definitions is misleading.
One should be aware that in typical circumstances ${\cal H}_0$
has an unbounded spectrum, unlike the perturbation ${\cal W}$.
Specifically we refer here to quantized chaotic systems.
In the standard representation ${\cal H}_0={\bf E}$ is a
diagonal matrix whose elements are the ordered energies $\{E_n\}$.
We limit our interest to a classically small (but quantum mechanically
large) energy interval where the mean level spacing $\Delta$ is
roughly constant. Assuming that the classical motion is characterized
by a {\em finite} correlation time $\tau_{cl}$, it follows that
the matrix ${\cal W}={\bf B}$ in the standard representation is
a banded matrix which {\em looks random} \cite{mario}.
Its bandwidth is $\Delta=2\pi\hbar/\tau_{cl}$.
The simplest model that captures this generic feature is
the Wigner banded random matrix model (BRM) \cite{wigner}.
The non-vanishing couplings are within the band $0 < |n-m| \le b$.
These coupling elements are zero on the average,
and they are characterized by the
variance $\sigma=(\langle |{\bf B}_{nm}|^2 \rangle)^{\rm {1/2}}$.
For sake of later convenience we define the bandwidth in energy units
as $\Delta_b=b\Delta$, and describe the band profile using a spectral
function $\tilde{C}(\omega)$
that equals $2\pi\hbar\sigma^2/\Delta$
for $|\hbar\omega|<\Delta_b$, and zero otherwise.
This is in fact the Fourier transform
of $C(\tau)=2b\sigma^2 \mbox{sin}(\tau/\tau_{cl})/(\tau/\tau_{cl})$.
The function $C(\tau)$ describes the time autocorrelation
of ${\cal W}$ in the Heisenberg picture.


We are interested in the time evolution of the energy distribution. The corresponding
time-dependent Schr\"odinger equation has been integrated numerically using the
self-expanding algorithm of \cite{wbr} to eliminate finite-size effects
(the remarkable accuracy is demonstrated in Fig.~1a).
In all the simulations the initial state $|\Psi_0\rangle$
is assumed to be an eigenstate $|m\rangle$ of ${\cal H}_0$.
In the first half period ($t<T/2$) we have ``conventional'' wavepacket
dynamics that we had studied in \cite{wbr}.
What we want to discuss in this Letter is the dynamics for $t>T/2$,
after the driving is reversed. The probability distribution after time $t$
is $P_t(n|m)=|\langle n|U(t)|m \rangle|^2$.
Averaging over realizations of the Hamiltonian
one obtains the average profile $P_t(n-m)$.
We characterize the evolving distribution
using two different measures: The survival probability
${\cal P}(t)={\cal P}(t;T)=P_t(r{=}0)$,
and the energy spreading
$\delta E(t)= \delta E(t;T)= \Delta \times \sqrt{\sum_r r^2 P_t(r)}$.
Thus the energy distribution at the end of the driving cycle is
characterized by ${\cal P}(T)={\cal P}(T;T)$ and $\delta E(T) = \delta E(T;T)$.

Looking on Fig.~1b and Fig.~1c we see that it is natural to define a time $t_r$ of
maximal return. This is the time, after the driving reversal, when we observe
maximum compensation.  The "compensation time" ($t_r$)  is determined either
by the minimum of the spreading $\delta E(t)$ or by the minimum of the
total transition probability $p(t)=(1-{\cal P}(t))$.
It is clear, by simple inspection of the plots in Fig.~1,
that the time of "maximal return" is in general not $t_r=T$
but rather $(T/2) \le t_r \le T$. Therefore we cannot regard
this compensation as an "echo" phenomena
(the term "echo" implies maximal return at $t_r=T$).
The dependence of $t_r$ on $T$ for various values
of $\epsilon$ is presented in Fig.2a and Fig.2b.


We want to understand the effect of driving reversal
on the spreading/decay processes, and in particular
to determine whether (or to what degree)
the spreading/decay processes are reversible.
The systematic strategy for the analysis of the dynamics
consists of two stages:
The first stage is to define the different $\epsilon$ regimes
in the corresponding parametric problem. As far as this
Letter is concerned \cite{rmrk} the important distinction
is between the standard perturbative regime ($\epsilon<\Delta/\sigma$),
the Wigner Lorentzian regime
($\Delta/\sigma < \epsilon < b^{1/2}\Delta/\sigma$),
and the non-perturbative semicircle regime
($\epsilon > b^{1/2}\Delta/\sigma$).
The second stage is to consider the time dependent
scenario separately in each regime.
The driving is characterized by two parameters:
One is the amplitude $\epsilon$,
and the other is the period $T$.
Thus we have a two dimensional parameter
space $(T,\epsilon)$ to study,
in analogy with having $(\Omega,A)$ space
in case of periodic driving \cite{rsp}.
If the perturbation parameter $\epsilon$ has the
meaning of "velocity", then the "standard perturbative regime"
is in fact the "adiabatic regime". By association
one naturally wonders whether there is also a "sudden regime".
We shall come to this issue later on.

It is natural  to inquire whether perturbation
theory is applicable for the analysis
of the driving reversal scenario.
Using first order perturbation theory one obtains
the following results for either the total transition
probability $p(t)=(1-{\cal P}(t))$
or the energy spreading $(\delta E(t))^2$:
\begin{eqnarray} \label{e4}
\epsilon^2 \times
\int_{-\infty}^{\infty}\frac{d\omega}{2\pi}
\left[\frac{1}{(\hbar\omega)^2}\right]
\tilde{F}_t(\omega) \tilde{C}(\omega)
\end{eqnarray}
where the expression in the square brackets should
be excluded in case of $\delta E(t)$.
The spectral content of the driving
is $\tilde{F}_t(\omega)=2(1-\cos(\omega t))$ for $t<T/2$,
and $\tilde{F}_t(\omega)=|1-2\mbox{e}^{i\omega T/2} +\mbox{e}^{i\omega t}|$
otherwise.
In Fig.~1b we display the rescaled energy
spreading $\delta E(t)/\epsilon$ for a
representative period $T$ and various $\epsilon$.
For small $\epsilon$ we observe the scaling which
is implied by Eq.(\ref{e4}). But for larger $\epsilon$
we observe a different (non-perturbative) behavior.
Similar results (Fig.~1c) are found in case
of $p(t)/\epsilon^2$.

Before we go on in explaining the new "physics"
in the non-perturbative regime, we would like to give,
for sake of completeness,  a very concise
summary of the "physics" in the perturbative regime.
The following is based on further analysis of Eq.(\ref{e4}),
extending the approach of Ref.\cite{wbr}.
In the "standard perturbative" (or "adiabatic") regime,
the spreading probability remains concentrated
all the time in the initial level.
In the "Wigner Lorentzian regime" the energy
spreading profile is characterized by two energy
scales $\Delta_b$ and $\Gamma=2\pi(\sigma\epsilon)^2/\Delta$.
If we are deep in this regime we have $1\ll\Gamma\ll\Delta_b$.
The energy scale $\Gamma$ is important for ${\cal P}(t)$,
because this is the energy range where most of the
probability is concentrated (the "core" component).
In contrast to that the energy scale that is important
for $\delta E(t)$ is $\Delta_b$, because the variance
is determined by the perturbative tails of the energy distribution,
and not by the width of the non-perturbative "core" component.
This means that in the Wigner Lorentzian regime the "physics" of
reversibility is different, depending whether we look on ${\cal P}(t)$
or on $\delta E(t)$. To be specific, for $\delta E(t)$ we have
reversibility provided $t<\tau_{cl}$, while for ${\cal P}(t)$
we have reversibility for $t<\hbar/\Gamma$.

We turn now to discuss the dynamics in the non-perturbative
regime, which is our main interest. In the absence of driving
reversal \cite{wbr} we get diffusion (\mbox{$\delta E(t)\propto \sqrt{t}$})
for $t>t_{\rm {prt}}(\epsilon)$, where
\begin{eqnarray}
\label{e5}
t_{\rm {prt}}(\epsilon) = \hbar/(\sqrt{b}\sigma\epsilon)\quad .
\end{eqnarray}
If $T/2<t_{\rm {prt}}(\epsilon)$, this non-perturbative diffusion
does not have a chance to develop, and therefore we can still
trust Eq.~(\ref{e4}). So the interesting case
is $T/2 > t_{\rm {prt}}(\epsilon)$,
which means large enough $\epsilon$.
It is extremely important to realize that without reversing the driving,
the presence or the absence of ${\bf E}$ cannot be detected.
It is only by driving reversal that we can easily determine (as in Fig.~1a)
whether the diffusion process is reversible or irreversible.

In order to analyze irreversibility we should distinguish
between two stages in the non-perturbative diffusion process.
The first stage ($t_{\rm {prt}}<t<t_{\rm {sdn}}$)
is reversible, while the second stage ($t>t_{\rm {sdn}}$) is irreversible.
[For much longer time scales we have recurrences or localization which are
not the issue of this Letter \cite{rmrk}.]
The second time scale ($t_{\rm {sdn}}$) is "new".
It did not appear in our "wavepacket dynamics" study \cite{wbr},
because it can be detected only by time/driving reversal experiment.

We can determine the time scale $t_{\rm {sdn}}$ as follows.
The diffusion coefficient is
$D_ {\rm {E}}=\Delta^2 b^{\rm {5/2}} \sigma\epsilon/\hbar$
up to a numerical prefactor \cite{wbr}.
The diffusion law is $\delta E^2(t)=D_{\rm {E}}t$.
The diffusion process is reversible as long as
${\bf E}$ does not affects the relative phases of
the participating energy levels. This means that
the condition for reversibility is
$(\delta E(t) \times t) / \hbar \ll 1$.
The latter inequality can be written
as $t< t_{\rm {sdn}}(\epsilon)$, where
\begin{equation}
\label{e6}
t_{\rm {sdn}}(\epsilon) = \left(\frac{\hbar^2}{D_ {\rm {E}}}\right)^{1/3} =
\left(\frac{\hbar^3}{\Delta^2b^{5/2}\sigma\epsilon}\right)^{1/3}\quad .
\end{equation}
This considerations imply that in the non-perturbative
regime we should have non-universal scaling with respect
to $\epsilon^{1/3}\times T$. This obviously
goes beyond any implications of perturbation
theory (Eq.(\ref{e4})). The numerical analysis (Fig.~2)
confirms the above claim.

The non-perturbative diffusion process involves only
one energy scale, and therefore in the non-perturbative regime
the scaling of $\delta E(T)$ and ${\cal P}(T)$ is similar (Fig.~3).
The non universal scaling of ${\cal P}(T)$
versus $\epsilon^{1/3}\times T$ is very different from
the observed behavior in the Wigner regime (not shown)
where the scaling is versus $\epsilon^2\times T$.
Also for $\delta E(T)$ we observe the same non-perturbative
scaling by plotting $\delta E(T)/\epsilon^{1/3}$
versus $\epsilon^{1/3}\times T$. The latter should
be contrasted with the perturbative $\epsilon^0\times T$
scaling which is observed in the Wigner regime (not shown).
The scaling of the vertical axis is explained
on the basis of diffusion process after $t_r$,
which is of the same nature as for $t<T/2$.

We are now able to identify the "sudden regime".
This is the $(T,\epsilon)$ regime which is defined
by the condition $T \ll t_{\rm {sdn}}(\epsilon)$.
The condition is never satisfied in the $\epsilon\rightarrow\infty$ limit.
If we interpret $\epsilon$ as "velocity" then this looks quite
strange in first sight, because we are used to associate the sudden limit
with having a very "fast" process. The subtlety is resolved
once we realize that in a proper textbook formulation of the
sudden limit we should take $\epsilon\rightarrow\infty$,
while holding the "displacement" ($\epsilon \times T$)
equal to a constant value.

In conclusion of the analysis of non-perturbative
features of QI in the PMP sense for Wigner model,
we would like to emphasize the distinction between
"reversible" and "irreversible" diffusion, and
the associated identification of the "sudden regime".



\centerline{\epsfig{figure=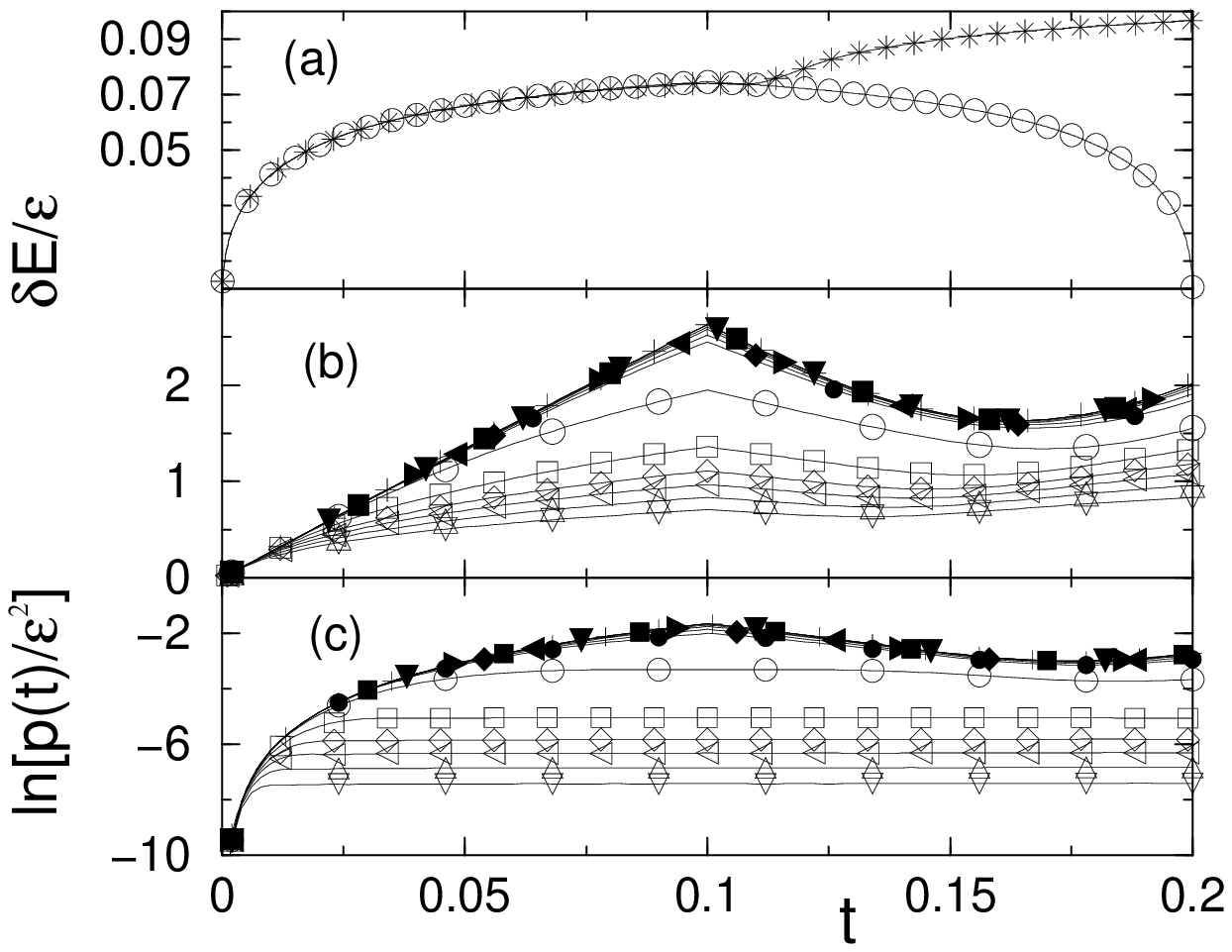,width=\hsize}}
{\footnotesize FIG.1:
Energy spreading and total transition probability as a function of time. We use
units such that $\Delta=\sigma=\hbar=1$, and $b=10$. Typically we take an average
over more than 100 different realizations in each run. (a) The actual evolution
($\star$) with ${\bf E}\ne0$ is compared with the evolution that would be obtained
($\circ$) if we had ${\bf E}=0$. The simulation is done for an extremely large
perturbation ($\epsilon=1000$). The remarkable reversibility in the ${\bf E}=0$
case is an indication for the accuracy of the numerical simulations. (b) The rescaled
energy spreading $\delta E(t)/\epsilon$, for $T=0.2$, and for $\epsilon\le 40$.
The clustering of curves corresponds to $\epsilon\le 2.5$. (c) The same for the
rescaled total transition probability $p(t)/\epsilon^2$.
}

\centerline{\epsfig{figure=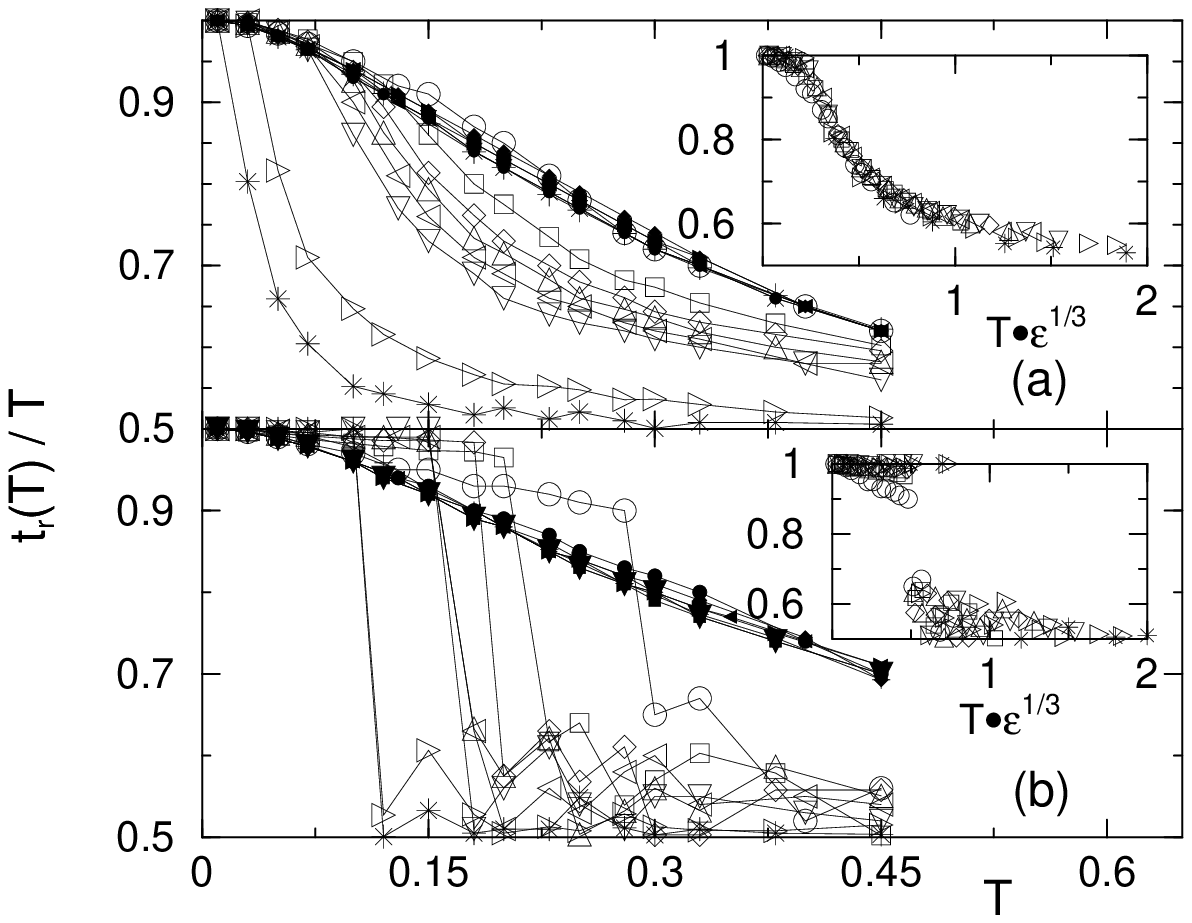,width=0.79\hsize}}
{\footnotesize FIG.2:
(a) The time $t_r$ calculated from $\delta E(t;T)$, as a function of
$T$ for amplitudes $\epsilon \le 40$. The clustering of curves corresponds to $\epsilon
\le 2.5$. The inset (where only data with $\epsilon>5$ are displayed) demonstrates
the scaling against $\epsilon^{1/3}\times T$ in the non-perturbative regime. (b) The
same, but now $t_r$ is evaluated from ${\cal P}(t;T)$.
}

\vspace*{0.5cm}

\centerline{\epsfig{figure=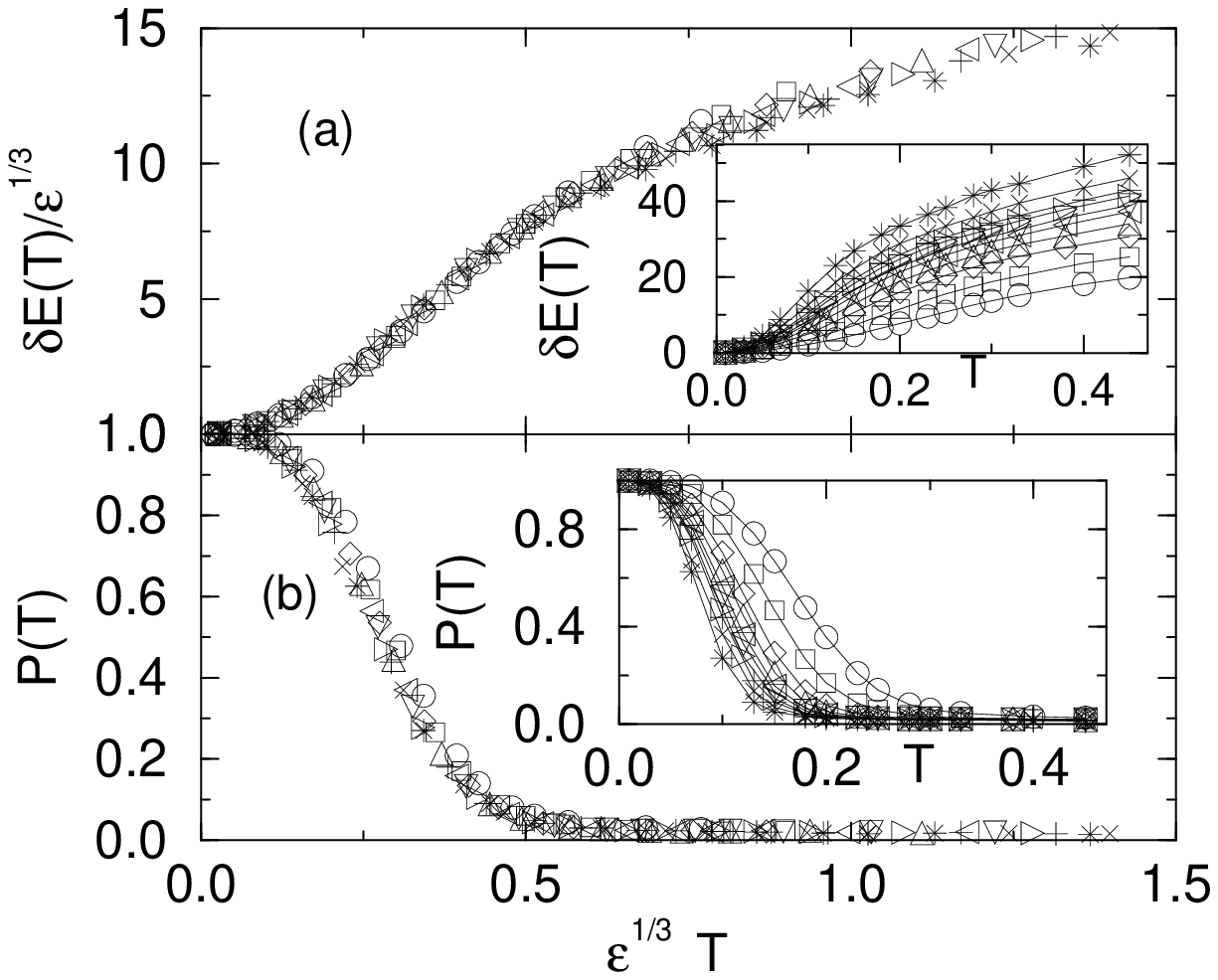,width=0.79\hsize}}
{\footnotesize FIG.3:
(a) The rescaled energy spreading $\delta E(T)/ \epsilon^{1/3}$
versus the rescaled time $\epsilon^{1/3}\times T$
for $\epsilon$~values in the non-perturbative regime ($\epsilon >5$).
The inset displays the same data without rescaling.
(b) The same for the survival probability ${\cal P}(T)$.
}


\begin{thebibliography}{0}

\bibitem{peres} A. Peres, Phys. Rev. A {\bf 30}, 1610 (1984).

\bibitem{quantcomp}
R.A. Jalabert and H.M. Pastawski, Phys. Rev. Lett. {\bf 86}, 2490 (2001);
F.M. Cucchietti, H.M. Pastawski, R. Jalabert Physica A, {\bf 283}, 285 (2000);
F.M. Cucchietti, H.M. Pastawski and D.A. Wisniacki, Phys. Rev. E {\bf 65}, 045206 (2002);
Ph. Jacquod, P.G. Silvestrov and C.W.J. Beenakker, Phys. Rev. E {\bf 64}, 055203 (2001);
N.R. Cerruti and S. Tomsovic, Phys. Rev. Lett. {\bf 88}, 054103 (2002);
T.~Prosen, quant-ph/0106149;
Z.P. Karkuszewski, C. Jarzynski, and W.H. Zurek, Phys. Rev. Lett. {\bf 89}, 170405 (2002).
D.A. Wisniacki and D. Cohen, Phys. Rev. E {\bf 66}, 046209 (2002).

\bibitem{wld} D. Cohen, Phys. Rev. E {\bf 65}, 026218 (2002).
This publication is regarding dephasing due to the
interaction with chaotic degrees of freedom. The relation
to the QI studies of Ref.\cite{quantcomp}  is explained there.
Note that main stream literature is dealing mainly with dephasing due to the
interaction with {\em bath} degrees of freedom. See for example
{\em Quantum Dissipative Systems}, U. Weiss, World Scientific, Singapore (1999);
{\em Quantum Transport and Dissipation}, T. Dittrich, P. Hänggi, G. L. Ingold, B. Kramer,
G. Schön, and W. Zwerger, Wiley-VCH, Weinheim (1998);
W. Zurek, Phys. Scripta {\bf 76}, 186 (1998).
Interaction with RMT bath is discussed by
P. A. Bulgac, G.D. Dang and D. Kusnezov, Phys. Rev. E {\bf 58}, 196 (1998).
Some recent related publications are:
D. Braun, F. Haake, and W.T. Strunz, Phys. Rev. Lett. {\bf 86}, 2913 (2001);
T. Gorin, T.H. Seligman, J. of Optics B {\bf 4}, S386 (2002).

\bibitem{rsp} D. Cohen and T. Kottos, Phys. Rev. Lett. {\bf 85}, 4839 (2000).

\bibitem{vrn} D. Cohen, Annals of Physics {\bf 283}, 175 (2000); D. Cohen in
"Proceedings of the International School of Physics Enrico Fermi Course CXLIII",
Edited by G. Casati, I. Guarneri and U. Smilansky, IOS Press, Amsterdam (2000).
Y.V.Fyodorov and A.D.Mirlin, Phys. Rev. Lett. {\bf 67}, 2405 (1991).

\bibitem{wigner} E. Wigner, Ann. Math {\bf 62} 548 (1955); {\bf 65} 203 (1957);
V.V. Flambaum, A.A. Gribakina, G.F. Gribakin and M.G. Kozlov, Phys. Rev. A {\bf 50}
267 (1994); G. Casati, B.V. Chirikov, I. Guarneri, F.M. Izrailev, Phys. Rev. E
{\bf 48}, R1613 (1993); \ Phys. Lett. A {\bf 223}, 430 (1996); Ph. Jacquod and D.
L. Shepelyansky, Phys. Rev. Lett. {\bf 75}, 3501 (1995); Y. V. Fyodorov, O. A.
Chubykalo, F. M. Izrailev, and G. Casati, ibid. {\bf 76}, 1603 (1996).


\bibitem{mario} M. Feingold and A. Peres, Phys. Rev. A {\bf 34} 591, (1986);
M. Feingold, D. Leitner, M. Wilkinson, Phys. Rev. Lett. {\bf 66}, 986 (1991);
M. Feingold, A. Gioletta, F. M. Izrailev, L. Molinari, ibid. {\bf 70}, 2936 (1993);
M. Wilkinson, M. Feingold, D. Leitner, J. Phys. A {\bf 24}, 175 (1991).

\bibitem{wbr}
D. Cohen, F. Izrailev and T. Kottos, Phys. Rev. Lett., {\bf 84} 2052 (2000);
T. Kottos and D. Cohen, Phys. Rev. E {\bf 64}, R-065202 (2001).

\bibitem{rmrk}
Long time recurrences are not an issue of the present Letter. This is the reason
why we do not make a distinction between the "ergodic" and the "localization"
regimes. See \cite{wbr} for details.


\end{thebibliography}
\end{document}